%% file: Spaper.tex
\documentclass[11pt]{article}

\usepackage{amsmath}
\usepackage{verbatim}
\usepackage{amsfonts}
\usepackage{latexsym}
\usepackage{epsfig}
\usepackage{color}
\if@twoside\oddsidemargin25mm
\else
\input{texdefs}

\begin{document}

\input{Stitle}

\setcounter{page}{1}

\input{Sintro}

\input{Sstring}

\input{Scalc}

\input{SintTadp}

\input{StwProp}

\input{SconStates}

\input{Sconcl}

\appendix 
\def\theequation{\thesection.\arabic{equation}}

\setcounter{equation}{0}
\input{Swstorus}

\setcounter{equation}{0}
\input{Stheta}

\bibliographystyle{paper}
{\small
\bibliography{paper}
}

\end{document}

%% file: texdefs.tex
\renewcommand{\d}{\mathrm{d}}

\DeclareMathSymbol{\mg}{\mathrel}{symbols}{"1D}

\newcommand{\ga}{\alpha}
\newcommand{\gb}{\beta}

\newcommand{\gd}{\delta}

\newcommand{\gf}{\phi}

\newcommand{\gm}{\mu}
\newcommand{\gn}{\nu}

\newcommand{\gl}{\lambda}
\newcommand{\gr}{\rho}

\newcommand{\gvth}{\vartheta}
\newcommand{\gs}{\sigma}
\newcommand{\gt}{\tau}

\newcommand{\gz}{\zeta}
\newcommand{\gp}{\pi}
\newcommand{\gps}{\psi}
\newcommand{\get}{\eta}

\newcommand{\gD}{\Delta}
\newcommand{\gF}{\Phi}

\newcommand{\cD}{{\cal D}}

\newcommand{\cF}{{\cal F}}
\newcommand{\cG}{{\cal G}}

\newcommand{\ui}{{\underline i}}
\newcommand{\uj}{{\underline j}}

\newcommand{\ra}{\rightarrow}

\newcommand{\der}{\partial}

\newcommand{\inv}{^{-1}}
\newcommand{\dsp}{\displaystyle}

\newcommand{\labl}[1]{\label{#1}}
\newcommand{\beq}{\begin{equation}}
\newcommand{\eeq}{\end{equation}}
\newcommand{\barr}{\begin{array}}
\newcommand{\earr}{\end{array}}
\newcommand{\equ}[1]{\begin{gather} #1 \end{gather}}

\newcommand{\arry}[2]{\begin{array}{#1} #2 \end{array}}

\newcommand{\pmtrx}[1]{\begin{pmatrix} #1 \end{pmatrix}}

\newcommand{\non}{\nonumber}
\newcounter{oldcounter}

\newcommand{\bder}{\bar\partial}
\newcommand{\baa}{{\bar a}}

\newcommand{\bc}{{\bar c}}

\newcommand{\bX}{{\bar X}}

\newcommand{\bgm}{{\bar\mu}}

\newcommand{\bgl}{{\bar\lambda}}

\newcommand{\bgs}{{\bar\sigma}}
\newcommand{\bgt}{{\bar\tau}}

\newcommand{\tgm}{{\tilde \mu}}

\newcommand{\tget}{{\tilde \eta}}

\newcommand{\tgD}{{\tilde \Delta}}

\newcommand{\Intr}{\mathbb{Z}}
\newcommand{\Cplx}{\mathbb{C}}

\newcommand{\ba}[2]{\[\begin{array}{#2}\label{#1}}
\newcommand{\ea}{\end{array}\]}
\newcommand{\be}{\begin{equation}}
\newcommand{\ee}{\end{equation}}
\newcommand{\bea}{\begin{eqnarray}}
\newcommand{\eea}{\end{eqnarray}}

\newcommand{\E}[1]{\mathrm{E_{#1}}}
\newcommand{\U}[1]{\mathrm{U(#1)}}
\newcommand{\SU}[1]{\mathrm{SU(#1)}}
\newcommand{\SO}[1]{\mathrm{SO(#1)}}

\newcommand{\brkt}[2]{\bigl[ ^{#1}_{#2} \bigr]}

%% file: Stitle.tex
\thispagestyle{empty}

\begin{flushright}
FTPI-MINN-03/29 \\
UMN-TH-2218/03 \\
hep-th/0311013
\end{flushright}
\vskip 2 cm
\begin{center}
{\Large {\bf 
Stringy profiles of gauge field tadpoles near orbifold singularities: 
\\[2ex]  
I.\ heterotic string calculations
}
}
\\[0pt]
\vspace{1.23cm}
{\large
{\bf Stefan Groot Nibbelink$^{a,}$\footnote{
{{ {\ {\ {\ E-mail: nibbelin@hep.umn.edu}}}}}}}, 
{\bf Mark Laidlaw$^{b,}$\footnote{
{{ {\ {\ {\ E-mail: mlaidlaw@perimeterinstitute.ca}}}}}}},
\bigskip }\\[0pt]
\vspace{0.23cm}
${}^a$ {\it 
William I. Fine Theoretical Physics Institute, 
School of Physics \& Astronomy, \\
University of Minnesota, 116 Church Street S.E., 
Minneapolis, MN 55455, USA. \\}
\vspace{0.23cm}
${}^b$ {\it  
Perimeter Institute for Theoretical Physics, \\
35 King Street North, Waterloo, Ontario N2J 2W9, Canada
}
\bigskip
\vspace{1.4cm} 
\end{center}
\subsection*{\centering Abstract}

Closed string theories on orbifolds contain both untwisted and twisted 
states. The latter are normally assumed to live exactly at the 
orbifold fixed points.  We perform a calculation of a gauge field tadpole 
amplitude and show that off--shell both the twisted and untwisted states 
give rise to non--trivial momentum profiles over the orbifold
$\Cplx^3/\Intr_3$. These profiles take the form of Gaussian
distributions integrated over the fundamental domain of the modular
parameter of the torus.  The propagators of the internal coordinate
fields on the torus world sheet determine the width of the
Gaussian profiles. These propagators are determined up to a single
normal ordering constant which must be bounded below to allow the
existence of the coordinate space representation of these Gaussians. 
Apart from the expected massless states, massive and even
tachyonic string excitations contribute to the profiles in some anomalous
$\U{1}$ models. However, when a tadpole is integrated over the internal
dimensions, these tachyonic contributions cancel in a non--trivial
manner.

\newpage

%% file: Sintro.tex
\section{Introduction}
\labl{sc:intro}

In this series of two papers we investigate some local properties of
string theory on orbifold singularities and their interpretation in
field theory. Due to pioneering investigations 
\cite{dixon_85,Dixon:1986jc,Dixon:1987qv} of strings on
orbifolds, it is  
commonly accepted that strings can propagate without difficulty on a
target space with orbifold singularities. This has been confirmed by
many investigations of zero mode properties of various types of string
theories on different orbifold background. Field theories on 
orbifolds always seem to require more care, particularly when
gravitational effects are involved. Close to the orbifold singularity,
strong curvature effects become important and can lead to a
breakdown of effective field theory methods. We therefore expect
that studying strings in the background of an orbifold singularity
might teach us something about how we should treat those singularities
in field theory.

For both field and string theoretical reasons we employ gauge field
tadpoles as local probes of orbifold singularities in these two
papers. Field theory arguments suggest that a one--loop
Fayet--Iliopoulos $D$--term tadpole
\cite{Fayet:1974jb,Fischler:1981zk} will be generated in heterotic 
string models with an anomalous $\U{1}$ \cite{Dine:1987xk}. 
This was soon confirmed by direct string calculations
\cite{Atick:1987gy,Dine:1987gj}. (A similar investigation has been
performed for orbifolded type I string theories
\cite{Poppitz:1998dj}.) It was recently realized that in 
five dimensional field theory models on $S^1/\Intr_2$
\cite{Ghilencea:2001bw,Barbieri:2001cz,Scrucca:2001eb,GrootNibbelink:2002wv,GrootNibbelink:2002qp},
higher dimensional field theory models 
\cite{vonGersdorff:2002us,Csaki:2002ur}, as well as in heterotic
string inspired models \cite{GrootNibbelink:2003gb} that 
these tadpoles are generated locally at orbifold fixed points. 
By supersymmetry these $D$--term tadpoles are accompanied by tadpoles 
for the internal $\U{1}$ gauge field strength at those
singularities. (In the case of five dimensional models this reduces to
a tadpole for the derivative of the real scalar of the vector multiplet.) 
In this paper we propose to verify the existence of these gauge field
tadpoles in string theory, and to investigate their properties.

In field theory it has been shown that these gauge tadpoles can lead
to (strong) localization effects over the extra dimensions 
\cite{GrootNibbelink:2002wv,GrootNibbelink:2002qp,Barbieri:2002ic,
Marti:2002ar}, which can be important for phenomenological
applications of higher dimensional models. The physical interpretation
of these field theory models is rather involved, since these tadpoles are
proportional to (derivatives of) delta functions at the fixed
points. The energies needed to resolve such singularities are much
higher than the assumed validity of the effective field theory. 
However, as string theory is assumed to be a complete theory, it
should solve this problem by introducing some sort of ultra--violet cut--off. 
In addition, the tadpole contributions due to bulk and fixed point
fields seem quite different in (string--inspired) field theory: The
former have profiles depending on the internal momenta, and therefore
extend into the extra  dimensions, while the latter are assumed to be
confined exactly at the orbifold singularities. As the corresponding
untwisted and twisted states in string theory are related to each
other by modular transformations, it is interesting to see to what
extent these differences persist.

We have decided to divide the presentation of our results into two 
separate publications. This paper focuses on the details of the string
calculation that establishes that the gauge field tadpole discussed in
\cite{GrootNibbelink:2003gb} also arises in string theory. 
As the physical interpretations of these results are of 
interest to a wider audience of both string and field theorists, we
postpone our detailed comparison between the string and field theory
results to the accompanying paper \cite{GNL_II}.

We have restricted ourselves to the well--studied heterotic
$\E{8}\times \E{8}'$ string on the non--compact six dimensional
orbifold $\Cplx^3/\Intr_3$. This choice is motivated by a number of
factors:  It is well--know that string models with an anomalous $\U{1}$ in 
their zero mode spectrum exist on $\Intr_3$ orbifolds 
\cite{Kobayashi:1997pb}.  As discussed 
above, on field theoretical grounds we would expect that tadpoles for
internal gauge fields are also generated within the string setting. By
considering this simple non--compact orbifold we avoid the additional
complications of winding modes and Wilson lines. A compact orbifold
would introduce at least one more scale (its volume) and this would
make it more difficult to trace how string theory introduces a
regularization of the delta--like singularities. In addition, there
has been some recent interest in the heterotic string, as the authors
in ref.\ \cite{Gukov:2003cy} find that it is possible to stabilize moduli in
the heterotic string compactifications on Calabi--Yau threefolds,
another way of regularizing singularities coming from some orbifolds.

The tadpole profile over the orbifold  will suggest that the parallels
between string and field theory can even be extended to off--shell
amplitudes. Strictly speaking string amplitudes are only defined
on--shell, but as off--shell amplitudes in field theory contain a
wealth of extra information, many authors have pursued the development
of off--shell string theory 
\cite{Cohen:1987pv,Samuel:1988ua,Samuel:1988uu,Lechtenfeld:1988th,Ukegawa:1991wz,Witten:1993cr,Li:1993za}. 
This progress was partly stimulated by applying stringy techniques in
field theory calculations 
\cite{Bern:1990fu,DiVecchia:1996uq}.
Moreover, to understand the dynamics of tachyons in unstable brane
configurations, an off--shell description is essential 
\cite{Belopolsky:1995sk,Ellwood:2003xc}. (Tachyons may also arise in
closed string theory \cite{Adams:2001sv}.) 
Off--shell amplitudes become dependent normal
ordering constants. These constants can be interpreted as coefficients
of conformal maps of the worldsheet torus to itself 
\cite{Bern:1990fu,Cuomo:2000de}. 
From the very outset our investigation had a different objective than
those works: They aim to describe the string (zero) modes off--shell,
while we are interested in the momentum dependence in the extra
dimensions. For a tadpole on an $\Intr_3$ orbifold, no four
dimensional momenta can flow in or out, therefore probing this
dependence necessarily requires going off--shell.

This paper is organized as follows: In section \ref{sc:string} we
define our notation for the heterotic string on the non--compact orbifold
$\Cplx^3/\Intr_3$. In section \ref{sc:calc} we give the string 
calculation of the gauge tadpole at one--loop in string 
perturbation theory. This calculation identifies the twisted
propagators as the characteristic widths of the (off--shell) momentum 
profiles for the various states on the orbifold. 
The twisted propagators are determined up to a single normal ordering
constant in section \ref{sc:twProp}. The properties of the twisted
propagators lead to a natural classification of inequivalent
sectors which give different contribution to the profiles of the local
gauge field tadpoles. In section \ref{sc:conStates} we investigate  
the various orbifold models to determine what type of states (tachyonic, 
massless or massive) contribute to these tadpoles within those
sectors. Our conclusions are summarized in section \ref{sc:concl}. In
appendix \ref{sc:wstorus} several useful properties of fermionic and
bosonic world sheet theories are reviewed, and  appendix
\ref{sc:theta} gives a brief overview of our conventions for the
elliptic functions that we use throughout the paper.


%% file: Sstring.tex
\section{Heterotic string on $\boldsymbol{\Cplx^3/\Intr_3}$}
\labl{sc:string}

In section \ref{sc:calc} we will perform a one--loop calculation in
heterotic string theory of gauge field tadpoles on the orbifold
$\Cplx^3/\Intr_3$; here we present the framework for that computation. 
We consider the world sheet torus parameterized by the
complex coordinate $\gs$ with Teichm\"uller
parameter $\gt$; the torus periodicities are defined as 
$\gs \sim \gs + 1$ and $\gs \sim \gs + \gt$. On this world sheet torus
the conformal field theories live in the heterotic string in
light--cone gauge: 
$X^M(\gs), M = 2,\ldots 9$ are the coordinate fields and $\gps^M(\gs)$
their right--moving fermionic partners. The part of the theory
encoding the gauge structure is described by the left--moving fermions 
$\gl^{2I}_a(\gs), \gl^{2I+1}_a(\gs)$. 
For the $\E{8}\times \E{8}'$ theory there are two sets,  labeled by $a =
1,2$, of  $I = 1,\dots 8$ fermions.  (The $\SO{32}$ string
contains $I= 1, \dots 16$ fermions and $a = 1$. As most of our
description applies to both theories, we can use both indices $I$ and
$a$ to describe the $\SO{32}$ and $\E{8}\times \E{8}'$ theory
simultaneously.)  For strings on the orbifold $\Cplx^3/\Intr_3$  it
is convenient to define complex combinations of these fields 
\equ{
X_{i} = \frac 1{\sqrt 2} ( X_{2i+2} + i X_{2i+3} ), 
\qquad 
\gps_i = \frac 1{\sqrt 2} ( \gps_{2i+2} + i \gps_{2i+3} ), 
\qquad
\gl^I_a =  \frac 1{\sqrt 2} ( \gl^{2I}_a + i \gl^{2I+1}_a )
}
and their complex conjugates for $i = 0, \dots 3$ in light--cone
gauge. The boundary conditions of these fields on the world sheet
torus are given by  
\equ{
\arry{ccc}{ \dsp 
X_i(\gs+1) = e^{-2\pi i \, p \gf_i} X_i(\gs), & & 
X_i(\gs+\gt) = e^{+2\pi i \, p' \gf_i} X_i(\gs),
\\[2ex] \dsp 
\gps_i(\gs+1) = e^{-2\pi i (p \gf_i +s/2)} \gps_i(\gs), & & 
\gps_i(\gs+\gt) = e^{+2\pi i (p' \gf_i +s'/2)} \gps_i(\gs),
\\[2ex] \dsp 
\gl_a^I(\gs+1) = e^{-2\pi i (p v^I_a +t_a/2)} \gl_a^I(\gs), & & 
\gl_a^I(\gs+\gt) = e^{+2\pi i (p' v^I_a +t_a'/2)} \gl_a^I(\gs),
}
\labl{BoundCond}
}
where $p, p' = 0,1,2$ label the different orbifold boundary conditions,
and $s, s', t_a, t_a' = 0,1$ define the different spin structures for
$\gps_i$ and $\gl^I_a$, respectively. The boundary conditions
\eqref{BoundCond} are fully specified by giving the spacetime and
gauge shifts $\gf_i$ and $v^I_a$, respectively. For the orbifold
$\Cplx^3/\Intr_3$ we have \cite{dixon_85,Dixon:1986jc} 
\equ{
3 \gf_i  = 0 \mod 1, 
\quad 
\frac 32 \sum_i \gf_i = 0 \mod 1, 
\qquad 
3 v_a^I = 0 \mod 1,
\quad 
\frac 32 \sum_{a,I} v^I_a = 0 \mod 1. 
}
Since we do not orbifold the four dimensional spacetime, 
we take $\gf_0 = 0$. By integral shifts and Weyl reflections, we can
always bring $\gf$ to the form $\gf = (0, 1, 1, \mbox{-}2)/3$, so that
$\sum_i \gf_i = 0$. Similarly, we will assume that gauge shifts have
been taken such that $\sum_{a,I} v_a^I = 0$. These constraints on the 
spacetime and gauge shifts
complete the definition of the different world sheet theories of the
orbifold model. We have collected many useful details of these twisted
world sheet theories in appendix \ref{sc:wstorus}.

Each different boundary condition, encoded by $p,p',\,s,s',\,t_a,t_a'$, 
defines a different world sheet theory. The full string theory is
obtained by combining all these possible boundary conditions. The
choices of relative phases between the sectors define GSO projections
\cite{Gliozzi:1977qd}, for the orbifold and three different spin 
structures \cite{Seiberg:1986by}.  One defines the full string
partition function as the sum   
\equ{
Z= \sum_{p,p', t_a, t_a'} Z{}^{p,p'}_{t_a, t_a'}, 
\qquad 
Z{}^{p,p'}_{t_a, t_a'} = 
 \sum_{s,s'} 
\tget^{p,p',\, s,s'}_{t_b, t_b'} \,
Z{}^{p,p', s,s'}_{t_a, t_a'}, 
\labl{WeightedPartFun} \\[2ex] \dsp 
Z{}^{p,p',s,s'}_{t_a, t_a'} =  \gm_0\,  
e^{ - 2 \gp i  \sum\limits_{a,I} v_a^I\, p \frac{t_a'}{2}}
\, 
\prod_{i = 0}^3 Z_{X}\brkt{p \,\gf_i}{p' \gf_i}(\gm_i | \gt)  \, 
Z_{\gps}\brkt{p\, \gf_i + \frac s2}{p' \gf_i+\frac{s'}2}(\gm_i | \gt) \,  
 \prod_{a, I}
Z_{\gl}\brkt{p \,v_a^I + \frac{t_a}{2}}{p' v_a^I + \frac{t_a'}2} (\gn^I_a | 
\gt)
\non 
} 
over the partition functions of the individual sectors.  These partition 
functions are given by 
\equ{
Z_{X}\brkt{\ga}{\gb}(\gm|\gt) =
\frac {\bigl|  \get(\gt) \bigr|^2} 
{ \bigl| \gvth\brkt{\frac 12 - \ga}{\frac 12 -\gb}(\gm|\gt)\bigr|^2}, 
\quad 
Z_{\gps}\brkt{\ga}{\gb}(\bgm|\bgt) = 
\frac{ \overline{\gvth\brkt{\frac 12- \ga}{\frac 12 - \gb}(\gm|\gt)} }
{ \overline{\get(\gt)}  }, 
\quad 
Z_{\gl}\brkt{\ga}{\gb}(\gn|\gt) = 
\frac{ \gvth\brkt{\frac 12- \ga}{\frac 12 - \gb}(\gn|\gt) }{ \get(\gt)  }, 
\labl{PartitionFun}
}  
where we have introduced the complex sources $\gm_i, \gn^I_a$ for
later convenience. Details can be found in appendix
\ref{sc:wstorus}. Defining the partition function as the limit of
$\gm, \gn \ra 0$ avoids having to treat the case $\ga=\gb = 0$
separately, provided that the total partition function is multiplied
by $\gm_0$. These partition functions have been computed in
\eqref{eq:freebospf} and   \eqref{eq:freefermpf}. (Since the fields
$\gps_i$ are right--movers we use the complex conjugate of the
left--moving result.) The phase factor in \eqref{WeightedPartFun} has
been introduced to ensure that the individual partition functions  
$Z{}^{p,p',s,s'}_{t_a, t_a'}$ be functions of the equivalence
classes of $p \sim p + 3$ and $s \sim s + 2$, $t_a \sim t_a +2$ 
\cite{Senda:1988wp,Senda:1988pf,Scrucca:2001ni}.

We impose the standard GSO projections for the
right--moving fermions $\gps$ and the left--moving fermions
$\gl_a$. In addition, we enforce a  generalized projection required by
the orbifold boundary conditions. The compatibility of these
projections is encoded in the factorization of the phases as 
\equ{
\get^{p,p',\, s,s'}_{t_a, t_a'} = \get^{s,s'} \tget^{p,p'}_{t_a,t_a'}, 
\quad 
\get^{s,s'} = \exp \{- \gp i\, s s' \}, 
\quad 
\get^{p,p'}_{t_a, t_a'} = \exp{ 2 \pi i \Bigl\{
\frac 12 \Bigl( \sum_i (\gf_i)^2 - \sum_{a,I}(v_a^I)^2 \Bigr) 
\, p p' 
\Bigr\} }. 
\labl{disentanglement}
} 
These phases are consequences of modular invariance, as been
investigated by various groups
\cite{Kawai:1986vd,Kawai:1987ah,Minahan:1988ha,Senda:1988pf,Senda:1988wp,Antoniadis:1988wp,Scrucca:2001ni}.
In our case we find that invariance under $\gt \ra \gt + 1$ and $\gt
\ra -1/\gt$ leads to the relations between the phases 
\equ{
\tget^{p,p'+p,\, s,s'+s}_{t_a, t_a'+t_a}  = 
e^{2\pi i \bigl\{ \sum\limits_{a,I}v_a^I p \frac {t_a}{2} + 
\frac 12 \sum\limits_i \bigl( p \gf_i + \frac {s}{2} \bigr)^2 -
\frac 12 \sum\limits_{a,I} \bigl(p v_a^I + \frac {t_a}{2}\bigr)^2 
\bigr\} }
\get^{p,p',\, s,s'}_{t_a, t_a'}, 
\\[2ex]
\tget^{\mbox{-}p',p,\, \mbox{-}s',s}_{\mbox{-}t_a', t_a}  = 
e^{ 2 \pi i \bigl\{  - \sum\limits_{a,I} v_a^I  
\bigl(p \frac{t_a'}{2} + p' \frac{t_a}{2}   \bigr) + 
\sum\limits_i \bigl(\frac 12 + p \gf_i + \frac {s}{2} \bigr) 
 \bigl(\frac 12 - p' \gf_i - \frac {s'}{2} \bigr) -
\sum\limits_{a,I} \bigl(\frac 12 + p v_a^I + \frac {t_a}{2}\bigr)
\bigl(\frac 12 - p' v_a^I - \frac {t_a'}{2}\bigr)
\bigr\} }
\get^{p,p',\, s,s'}_{t_a, t_a'}, 
\non 
}
respectively. The first terms in these exponents arise because of the
phase in the partition functions \eqref{WeightedPartFun}. 
Consistency of the solution \eqref{disentanglement} 
is ensured by requiring that the following
conditions on the spacetime and gauge shifts are fulfilled
\equ{
\frac 12 \sum_{a,I} v_a^I - \frac 12 \sum_i \gf_i = 0 
\mod 1,
\qquad 
\frac 32 \sum_{a, I} (v_a^I)^2 - 
\frac 32 \sum_i (\gf_i)^2 = 0 \mod 1. 
\labl{consistency} 
}
The second condition follows upon requiring modular invariance under
the transformation $\gt \ra \gt +3$ in the sector $s= t_a = 0$, see 
\cite{Vafa:1986wx,Ibanez:1988pj}.

The sum over the spin structures $s, s'$ can be removed by applying
the Riemann's identity \eqref{RiemannSimple}, and we find 
\equ{
Z^{p,p'}_{t_a, t_a'}(\gm, \gn|\gt) = 
\tget^{p,p'}_{t_a, t_a'}  
\frac {2 \gm_0}{ \get(\gt)^{12}} 
\prod_{a, I}  
\gvth\brkt{\frac 12 - p \, v_a^I - \frac{t_a}2}
{\frac 12 - p' v_a^I - \frac{t_a'}2}(\gn_a^I|\gt) 
\prod_{i=0}^3 
\frac{ 
\overline{
\gvth\brkt{\frac 12 - p\, \gf_i}{\frac 12 - p' \gf_i}
(\mbox{-}\tgm_i|\gt)
}}{
\Bigl|
\gvth\brkt{\frac 12 - p \, \gf_i}{\frac 12 - p' \gf_i}
(\, \gm_i|\gt)  
\Bigr|^2}, 
\labl{CharPart}
} 
with the definition
\equ{
\tilde \gm_i = - \gm_i + \frac{1}{2} \sum_j \gm_j.
}
Notice that in the limit of $\gm_i \ra 0$ the partition function
becomes a holomorphic function of $\gt$. (In fact --as is well--known--
the whole partition function is zero in this limit; precisely for that
reason we have introduced the $\gm_0$.) Hence the expression above
simplifies to  
\equ{
Z^{p,p'}_{t_a, t_a'}(\gn|\gt) = 
\frac {-1}{2\pi} 
\tget^{p,p'}_{t_a, t_a'} 
\, 
\frac 1{\get(\gt)^{15}} 
\prod_{a,I} \gvth\brkt
{\frac 12 - p\, v_a^I - \frac{t_a}{2}}
{\frac 12 - p' v_a^I - \frac{t_a'}{2}}(\gn|\gt)
\left(
\prod_{i=1}^3
\gvth\brkt{\frac 12 - p \,\gf_i}{\frac 12 - p' \gf_i}(0|\gt)
\right)\inv. 
\labl{HoloPartFun}
}


%% file: Scalc.tex
\section{String calculation of tadpoles of internal gauge fields}
\labl{sc:calc}

We investigate the local tadpole structure of Cartan gauge fields with
internal spacetime indices by calculating the expectation 
values of the normal ordered vertex operators 
\equ{
V_j^{bJ} =\
: \! ( \bder X_j + i \, k_M \gps^M \gps_j) \,
\bgl^{J}_b \gl^{J}_b \, e^{i k_M X^M} \! : 
\labl{vertexoperator}
}
at the one loop (torus) order.  The vertex operators $V_\uj^{bJ}$ are
defined as $V_j^{bJ}$ with the replacement $j \ra \uj$. As the
expectation values of these vertex operators are closely related, we
focus on one of them only, and make this relation explicit at the
point where this is most convenient. No sum over $b$ is implied in
\eqref{vertexoperator}, and the combination of gauge indexed fermions
$\bgl^{J}_b \gl^{J}_b$ has been chosen so that 
the vertex operator is in the Cartan subalgebra of the gauge group.
Since string theory expectation values are an average of free field theory 
expectation values, we calculate the expectation value of the vertex 
operator as a weighted average with respect to the partition functions 
\eqref{WeightedPartFun}. We find that 
\equ{ 
\langle V_j^{bJ} (k) \rangle 
=
i k_\uj \, \frac{\gd^4(k_4)}{(2\gp)^4}\, 
\langle G_j^{bJ}(k_6) \rangle 
\qquad 
\langle G_j^{bJ}(k_6)\rangle  = \sum Z{}^{p,p', s,s'}_{t_a, t_a'} 
\langle G_j^{bJ} (k_6)\rangle{}^{p,p', s,s'}_{t_a, t_a'},
\labl{ExpVert}
}
where the sub-- and superscripts on the expectation values denote
that they are evaluated within the corresponding set of boundary
conditions. The subscripts on $k_4$ and $k_6$ indicate that these
momentum vectors lie in the four dimensional Minkowski or the six
dimensional internal space, respectively, with $k = (k_4, k_6)$. 
The brackets $\langle \rangle$ without subscripts refer to
the sum of expectation values in the different sectors weighted by
the corresponding partition functions. The four dimensional delta
functions result from the zero mode integral of  $\exp({i k_\gm X^\gm})$
and imposes four dimensional momentum conservation.

The dependence on the bosonic fields $X_i$ is defined by a point
splitting regularization in the following manner:  We consider 
exponentials  
$\bder X_j \, \exp(i k_j X_\uj + i k_\uj X_j)$,
$\bder  X_\uj \, \exp(i k_j X_\uj + i k_\uj X_j)$ and 
$\exp(i k_j X_\uj + i k_\uj X_j)$
with the bosonic fields $X_j(\gs)$ and $X_\uj(\gs')$ evaluated at
different points $\gs \neq \gs'$ on the torus world sheet. Next we use
\eqref{ExpCorr} to express the expectation values of the exponentials
in terms of bosonic propagators defined in \eqref{eq:fbprop}, and
finally, we take the limit of zero separation.   
(One of the two derivatives comes with an opposite sign, as it is a
derivative with respect to the second argument of $\tgD_X$.)
In this way we find that in a particular sector
\equ{
\langle G_j^{bJ} \rangle{}^{p,p', s,s'}_{t_a, t_a'} = 
\Bigl( - \bder \gD_{X}\brkt{p \,\, \gf_j}{p' \gf_j}
+ \gD_{\gps} \brkt{p\,\, \gf_j+\frac s2}{p' \gf_j+ \frac {s'}2}
\Bigr)
\gD_\gl\brkt{p \,\,v^J_b+\frac {t_b}2}{p' v^b_J+ \frac {t_b'}2}
\,
\prod_{i=1}^3
e^{- k_\ui k_i \, \gD_X\brkt{p\,\, \gf_i}{p' \gf_i}}.
\labl{Gauss}
}
This is expressed in terms of the normal ordered propagators 
$\gD_X$, $\gD_\gl$ and $\gD_\gps$ at zero world--sheet separation.  
The fermionic propagators $\gD_\gl$ and $\gD_\gps$ are given by
\eqref{NormFermProp} and its complex conjugate.

We use conformal normal ordering to remove any singularities that may 
arise in this limit, and this introduces normal ordering constants for
the correlator of the bosonic fields.  
On--shell amplitudes do not depend on these arbitrary constants. Also
in the present case we see, that for the on--shell tadpole ($k_i = 0$)
the possible normal ordering constants drop out. However, as we are
interested in the off--shell properties of the tadpole, we
expect to find some dependence on these constants. We return to this
point at the end of section \ref{sc:twProp}.

Taking the limit of zero separation on the equation 
\equ{
 \bder \tgD_X\brkt{p\,\, \gf_i}{p' \gf_i} = 
\tgD_\gps\brkt{p\,\, \gf_i}{p' \gf_i}, 
} 
which can be deduced from \eqref{eq:ffprop} and \eqref{eq:fbprop}, 
we re--express \eqref{Gauss} as 
\equ{
\langle G_j^{bJ} \rangle{}^{p,p', s,s'}_{t_a, t_a'} =
\Bigl( - \gD_{\gps}\brkt{p \,\, \gf_j}{p' \gf_j}
+ \gD_{\gps} \brkt{p\,\, \gf_j+\frac s2}{p' \gf_j+ \frac {s'}2}
\Bigr)
\gD_\gl\brkt{p \,\,v^J_b+\frac {t_b}2}{p' v^b_J+ \frac {t_b'}2}
\,
\prod_{i=1}^3
e^{- k_\ui k_i \, \gD_X\brkt{p\,\, \gf_i}{p' \gf_i}}.
\labl{Gauss2}
}
Inserting the expressions for the normal ordered fermionic 
correlators \eqref{NormFermProp} for $\gD_\gl$ and the complex
conjugate for $\gD_\gps$ and combining this with the character valued
partition function \eqref{CharPart}, we find that
\equ{
\langle G_j^{bJ} \rangle =
\sum_{p,p',t_a, t_a'}
\left.
 \frac{\der}{\der \gn_b^J}
\frac{\der}{\der \bgm_j}
Z^{p,p'}_{t_a, t_a'}(\gm, \gn)
 \, 
\prod_{i=1}^3
e^{- \bar k_i k_i \, \gD_X\brkt{p\,\, \gf_i}{p' \gf_i}}
\right|_{\gm = \gn = 0}.
\labl{GaussFact}
}
To write this we have used that the differentiation with respect to 
$\bgm_j$ of the partition function $\smash{Z^{p,p'}_{t_b,t_b'}}$
precisely gives the $\gD_{\psi_j}$ correlators (with and without the
spin structures) of \eqref{Gauss2}. To evaluate the derivative with
respect to $\bgm_j$  in the limit of $\gm \ra 0$,  we only need to
consider the anti--holomorphic  part of \eqref{CharPart}, hence we obtain
\equ{
\bgm_0
\left. \frac{\der}{\der \bgm_j}
Z^{p,p'}_{t_a,t'_a}(\gm,\gn|\gt, \bgt)  \right|_{\gm=0}
 =
\frac 12 Z^{p,p'}_{t_a,t'_a}(\gn|\gt),
\labl{GetHolo}
}
where the holomorphic partition function $Z^{p,p'}_{t_a,t'_a}(\gn|\gt)$ is
given in \eqref{HoloPartFun}.The rescaling with $\bgm_0$ is required, 
otherwise the result diverges in the limit  $\gm \ra 0$. The local
gauge field tadpole then becomes a function of the holomorphic
partition function only, and hence only depends on the $N=1$
multiplets rather than their individual bosonic and fermionic
constituents.  This is a general feature of the application of the
Riemann's identity \eqref{RiemannSimple} within string theory.
Another way to express this is to say that the sum over the spin 
structures $s$ gives the sum over space--time bosons and fermions, and
thus we have a contribution from both bosons and fermions in what may be 
thought of as a field theory one--loop diagram.

The expectation values \eqref{ExpVert} of the vertex operators
$V_j^{bJ}$ and $V_\uj^{bJ}$ can be written as 
\equ{ 
\langle V_j^{bJ} (k) \rangle 
=
i k_\uj \,\frac{\gd^4(k_4)}{(2\gp)^4} \, 
\langle G^{bJ}(k_6) \rangle, 
\qquad
\langle V_\uj^{bJ} (k)\rangle =
- i k_j  \, \frac{\gd^4(k_4)}{(2\gp)^4}\, 
\langle G^{bJ}(k_6) \rangle.
}
Because of \eqref{GetHolo}, the functions $G^{bJ}$ do not depend on
the internal spacetime indices $j, \uj$. Moreover, the sign of the
expectation values of $V_j^{bJ}$ and $V_\uj^{bJ}$ are opposite for two
reasons: Since $\gps_j$ and $\gps_\uj$ are fermions, interchanging
their order in \eqref{vertexoperator} gives a relative minus
sign. Secondly, taking derivatives with respect $l$ and $\bar l$
of equation \eqref{ExpCorr} gives the expectation values of
$\bder X_\uj \exp( ik_i X_\ui + i k_\ui X_i)$ and
$\bder X_j \exp( ik_i X_\ui + i k_\ui X_i)$ with a relative sign again.

We can Fourier transform $G^{bJ}(k_6)$ 
over the orbifolded dimensions to obtain $\cG^{bJ}(z)$ 
and integrate the result over the fundamental domain
$\cF$ of the modular parameter $\gt$, defining
\equ{
\cG^{bJ} (z) =
\int_\cF \frac{\d^2 \gt}{\gt_2^2} \cG^{bJ} (z|\gt).
}
These considerations imply that the corresponding effective field theory 
interaction is given by 
\equ{
S_{FI} = \int \d^{10}x \, \sum_{j, b, J}
\bigl( \der_\uj A_j^{bJ} - \der_\uj A_j^{bJ} \bigl)
\cG^{bJ} (z).
\labl{LocalFI}
}
Since only the exponential
factor in \eqref{GaussFact} depends on the external six dimensional
momentum $k$, the Fourier transform is obtained easily
\equ{
\cG^{bJ}(z|\gt) =
\sum_{p, p',t_a, t_a'}
\left. \frac 12
\frac{\der}{\der \gn_b^J}
Z^{p,p'}_{t_a,t_a'}(\gn)
\right|_{0}
\prod_{i=1}^3 \frac {2\gp}{ \gD_X\brkt{p\,\, \gf_i}{p' \gf_i}}
e^{- \bar z z / \gD_X\brkt{p\,\, \gf_i}{p' \gf_i}}.
\labl{GaussSpatial}
}
This shows that $\gD_X(\gt)$ can be interpreted as the width of a 
complex three dimensional Gaussian distribution for a given value of the 
modular parameter $\gt$. (This is consistent with the observations 
in \cite{Kiritsis:1994jk,Kiritsis:1994ij,D'Appollonio:2003dr} that
wavefunctions of the gravitational wave states are Gaussians of widths 
specified by the orbifold twist.) The existence of the Fourier transform 
requires that $\gD_X(\gt) > 0$ for all $\gt$ in the fundamental domain
$\cF$. According to equation \eqref{GaussSpatial} the contributions to the
profile of the tadpole depend on the boundary conditions corresponding to
the distinct sectors of the orbifolding. This gives a measure to what 
extent the states of these different sectors are localized near the
fixed point of the orbifold $\Cplx^3/\Intr_3$. As this information is
encoded in  the functions $\gD_X\brkt{p\,\, \gf_i}{p' \gf_i}$ their
computation in section \ref{sc:twProp} is of central importance.

\subsection*{The integrated tadpole}

While the central theme of this work is the local structure of
tadpoles, we would like to make a couple of relevant comments about
the global properties. The first all, at the zero mode level the gauge field
tadpole vanishes trivially, since for a constant gauge field
background the field strength \eqref{LocalFI} is identically zero. 
However, the function multiplying the internal field strength
$F_{j \uj}^{bJ} = \der_j A_\uj^{bJ} - \der_\uj A_j^{bJ}$ in that
expression can be integrated over the full orbifold
\equ{
\int_{\Cplx^3/\Intr_3} \d^6 z\, \cG_j^{bJ}(z) =
\frac 13
\int_\cF \frac{\d^2 \gt}{\gt_2^2}
\left.
\frac 12
\frac{\der}{\der \gn_b^J}
Z(\gn|\gt)
\right|_{0}.
\labl{IntTadp}
}
This result follows immediately, since \eqref{GaussSpatial} contains 
properly normalized Gaussian distributions. In addition, the arguments
presented in ref.\ \cite{GrootNibbelink:2003gb}
lead us to expect that this integrated tadpole is proportional to the
zero mode $D$--term. Using the method of computing the integral over
the fundamental domain of a holomorphic function of $\gt$ explained in
ref.\ \cite{Lerche:1988qk}, it follows that only the massless string modes
contribute.\footnote{Unfortunately, as the local tadpole
  \eqref{GaussSpatial} is not holomorphic in $\gt$ because it depends
  on $\gD_X(\gt,\bgt)$, such powerful complex
  function techniques cannot be applied to our local results.} 
This means that the integrated tadpole is proportional to
the sum of $\U{1}$ charges of these zero modes. We will use this as a
cross check of our results for the local tadpoles in section
\ref{sc:conStates}. In this sense our calculations are a direct
extension of the results of Atick et al. \cite{Atick:1987gy}.

%% file: SintTadp.tex
%% file: StwProp.tex
\section{The twisted propagator}
\labl{sc:twProp}

In the previous section we found that the twisted propagators can be
interpreted as the width of Gaussians in momentum or coordinate space
characterizing the profiles of the gauge field tadpoles. Therefore, it is
important to determine them explicitly for the
orbifold$\Cplx^3/\Intr_3$. Since for this $\Intr_3$ 
orbifold we can choose $\gf= (1,1, \mbox{-}2)/3$, the three functions 
$\gD_X\brkt{p\,\, \gf_i}{p' \gf_i}$ for $i =1,2,3$ are the same; this
reflects the rotational symmetry of this orbifold. (This paper
specifically focus on the $\Intr_3$ orbifold, however, the method of
determining the relevant twisted propagators can easily be extended to
more general $\Intr_N$ orbifolds.)

The correlator of a boson with non--trivial boundary
conditions \eqref{boundBoson} with $\ga = p/3$, $\gb = p'/3$ and 
 $p, p' \in \left\{0,1,2\right\}$ and not both zero reads 
\equ{
\tgD_X\brkt{p/3}{p'\!/3}(\gs|\gt) 
= -
\frac 1{2\pi} \sum_{m,n} 
\frac { 2\gt_2}{|\gt(m+ p/3) + n + p'\!/3|^2} 
\gF\brkt{m+p/3}{n\, +p'\!/3}(- \gs| \gr).
\labl{correlatorBoson}
}
The mode functions $\gF\brkt{\ga}{\gb}$ are given in \eqref{modes} and
using those we can write the formal series expansion for this
correlator as  
\equ{
\tgD_X \brkt{p/3}{p'\!/3}(\gs|\gt) 
= -  
3^2 \, \frac 1{2\gp}\,  
\sum_{m,n} \frac{2 \gt_2
\exp\Bigl\{ - 2\pi i \, 
\frac { 
\gs' ( \bgt(3 m + p) + 3 n +p') 
- \bgs' ( \gt(3 m + p) + 3 n +p')
}
{\bgt - \gt}
\Bigr\}
}{|\gt(3m+ p) + 3n + p' |^2}, 
} 
where we have reparameterized $\gs = 3 \gs'$. We define the
projector 
\equ{
\gd_3(m) = \frac 13 \sum_{k=0}^2 e^{2\pi i\, km/3} = 
\begin{cases}
1, & m = 0 \mod 3,  \\ 
0, & m \neq 0 \mod 3,
\end{cases}
\labl{projectors}
}
and obtain 
\equ{
\tgD_X\brkt{p/3}{p'\!/3}(\gs|\gt) 
= - 
3^2 \, \frac 1{2\pi}\, 
\sum_{m',n'}' 
\gd_3(m'-p) \gd_3(n'-p') \, 
\frac{2 \gt_2
\exp\Bigl\{ - 2\pi i \, 
\frac { \gs' (\bgt m' + n' )- \bgs' ( \gt m' + n')} {\bgt - \gt}
\Bigr\}
}{|\gt m' + n' |^2}. 
} 
The restriction the sum without $(m',n')= (0,0)$, denoted by
the prime on the sum, is consistent by virtue of the assumption that
$p$ and $p'$ are not both zero modulo three. By inserting the definition 
\eqref{projectors} of the projectors the sums over $m', n'$ can be
cast in the form of the untwisted correlator \eqref{untCorr}, and we
find the twisted propagators can be written as sums over 
untwisted propagators
\equ{
\tgD_X\brkt{p/3}{p'\!/3}(\gs|\gt) = 
\sum_{k, l  = 0}^2 e^{- 2\pi i(p k + p' l)/3}\, 
\tgD \Bigl( \frac {\gs + k - l \gt }3 \Big| \gt  \Bigr).
\labl{TwProp}
}
The correlators in the  zero separation limit with all singular terms
removed are denoted by $\gD$ without the tilde. For all $(k,l) \neq 0$
the limit $\gs \ra 0$ does not lead to any singularity, and can be
taken readily. This leaves the case $k = l = 0$, but this one is
determined in \eqref{NormUnProp}. Therefore the expectation value the
twisted propagator at zero separation is given by  
\equ{
\gD_X\brkt{p/3}{p'\!/3}(\gt) = 
\sum_{(k, l) \neq 0} e^{- 2\pi i(p k + p' l)/3}\, 
\tgD \Bigl( \frac {k - \gt l}3 \Big| \gt \Bigr) 
- \ln (2 \gt_2) + \tilde c.
\labl{NormTwProp}
} 
The constant $\tilde c$ denotes the normal ordering constant for the
untwisted propagator. It is important to note that the dependence of
all normal ordered twisted propagators $\gD\brkt{p/3}{p'\!/3}$ on this
normal ordering constant is the same. 
Not all propagators for $p, p' = 0, 1,2$ are independent: 
Using the projector  \eqref{projectors} and the definition 
\eqref{correlatorBoson} in the $\gs \rightarrow 0$ limit it follows 
immediately
that  
\equ{
\gD_X\brkt{p/3+1}{p'\!/3}(\gt) = 
\gD_X\brkt{p/3}{p'\!/3+1}(\gt) = 
\gD_X\brkt{-p/3}{-p'\!/3}(\gt) = 
\Bigl( \gD_X\brkt{p/3}{p'\!/3}(\gt) \Bigr)^* =
\gD_X\brkt{p/3}{p'\!/3}(\gt). 
\labl{SymgDX}
}
Using \eqref{SymgDX} and the definition of
the projector \eqref{projectors} we obtain  
\equ{
\arry{ll}{
\gD_X\brkt{p/3}{p'\!/3}(\gt) &= 
(3 \gd_3(p) -1 ) \tgD(\mbox{$\frac 13$}|\gt) + 
(3 \gd_3(p') -1 ) \tgD(\mbox{$\frac \gt 3$}|\gt) + 
(3 \gd_3(p+p') -1 ) \tgD(\mbox{$\frac {\gt-1}3$}|\gt) + 
\\[2ex] 
&+\, (3 \gd_3(p-p') -1 ) \tgD(\mbox{$\frac {\gt+1}3$}|\gt)  
- \ln (2 \gt_2) + \tilde c.
}
\labl{twCorr}
}
The term $\ln (2 \gt_2)$ drops out of this expression all together,
using the expression for the untwisted propagator \eqref{UnProp}.

The full string amplitude is defined by an integral over the fundamental 
domain. As the fundamental domain is symmetric under 
$\gt_1 \ra -\gt_1$, it is important to know how the twisted
propagators transform under this reflection. An straightforward
analysis gives 
\equ{
\gD_X\brkt{p/3}{p'\!/3}(-\gt_1,\gt_2) = 
\gD_X\brkt{\mbox{-}p/3}{~p'\!/3}(\gt_1,\gt_2) = 
\gD_X\brkt{~p/3}{\mbox{-}p'\!/3}(\gt_1,\gt_2). 
\labl{tau1reflect}
}
This shows that the twisted correlators with $p=0$ or $q=0$ are even
under  $\gt_1 \ra - \gt_1$. From this discussion we conclude that
there are four different propagators: 
\equ{
\arry{c}{
\gD_u(\gt) = \gD_X\brkt{~0}{1/3}(\gt) = 
\gD_X\brkt{~0}{\mbox{-}1/3}(\gt), 
\qquad 
\gD_t(\gt) = \gD_X\brkt{1/3}{~0}(\gt) = 
\gD_X\brkt{\mbox{-}1/3}{0}(\gt), 
\\[2ex]
\gD_{d_+}(\gt) = \gD_X\brkt{1/3}{1/3}(\gt) 
= \gD_X\brkt{\mbox{-}1/3}{\mbox{-}1/3}(\gt),
\qquad 
\gD_{d_-}(\gt) = \gD_X\brkt{~1/3}{\mbox{-}1/3}(\gt) 
= \gD_X\brkt{\mbox{-}1/3}{~1/3}(\gt), 
}
}
which we will use to characterize the sectors they come from: untwisted 
($u$), twisted ($t$) and double twisted
($d_\pm$). It will be convenient to sometimes interpret these
symbols $u, t$ and $d_\pm$  as set of point $(p,p')$: 
$u = \{ (0, \frac 13), (0, \frac 23)\}$, 
$t = \{ (\frac 13,0), (\frac 23,0)\}$, 
$d_+ = \{ (\frac 13, \frac 13), (\frac 23, \frac 23)\}$, and 
$d_- = \{ (\frac 23, \frac 13), (\frac 13, \frac 23)\}$. 
The double twisted states have non--trivial periodicity conditions
around both cycles of the world sheet torus. Using the classification
we can write the expression for the tadpole profile as 
\equ{
G_j^{bJ}(k_6) = \int_\cF \frac {\d^2 \gt}{\gt_2^2} \, 
\sum_{s = u, t, d_\pm} \, 
Q_s^{bJ} \, 
e^{- \gD_s(\gt) k_i k_\ui}, 
\qquad 
Q_s^{bJ} = \sum_{(p, p') \in s}
\sum_{t_a, t_a'} 
\left. \frac 12
\frac{\der}{\der \gn_b^J} 
Z^{p,p'}_{t_a,t_a'}(\gn)
\right|_{0}.
}
Because the fundamental domain is invariant under 
$\gt_1 \ra -\gt_1$, the contributions of the sectors $d_+$ and $d_-$
to the full integral are the same, since from \eqref{tau1reflect} it
follows that 
\(
\gD_{d_+}(-\gt_1,\gt_2) = \gD_{d_-}(\gt_1,\gt_2).
\)

Another important thing we learn from this expression is that one
has to require that all propagators $\gD_s(\gt) > 0$ for all
$\gt\in\cF$: If there were to be a region of the fundamental domain
$\cF$ in which a propagator would be negative, it implies that 
the momentum profile function grows as a positive power of 
$\exp(k_i k_\ui)$, which is physically unacceptable because the 
Fourier transform to coordinate space does not exist. 
It can be shown that $\gD_{d_\pm}(\gt)$ takes the smallest value of all
propagators at the two end points of the fundamental domain 
$\gt_\pm = (\mp 1 + \sqrt 3\, i)/2$. This condition
leads to a lower bound on the normal ordering constant:
\equ{
\tilde c \geq \tilde c_{0} = 
\ln \Bigl| 
\frac 
{\gvth_1^2(\frac {\gt_+ - 1}{3}|\gt_+) \gvth_1'(0| \gt_+)}
{\gvth_1(\frac {\gt_+ + 1}{3}|\gt_+)  
\gvth_1(\frac {\gt_+}{3}|\gt_+)  
\gvth_1(\frac {1}{3}|\gt_+) }
\Bigr|^2. 
}
in terms of $\gt_-$ a similar expression can be given. 
We will argue in our next paper \cite{GNL_II}, which focusses
more on the phenomenological aspects of the tadpoles, that saturation
of the bound might be preferred.

%% file: SconStates.tex
\section{Model specific analysis}
\labl{sc:conStates}

Our analysis has been essentially model independent up to this
point. However, as has been investigated at length using field theory
methods \cite{GrootNibbelink:2003gb,Gmeiner:2002es} the local 
tadpoles associated with (anomalous) $\U{1}$'s depend very
sensitively upon the particular model examined. In table
\ref{tb:Z3shifts} we have summarized the five possible
$\Cplx^3/\Intr_3$ models within the heterotic $\E{8}\times \E{8}'$
string theory by giving their defining gauge shifts $v_a$ and the
unbroken gauge group in the effective four dimensional field theory of
string zero modes. These gauge shifts are uniquely defined up to
$\E{8}\times \E{8}'$ root lattice shifts and Weyl reflections, which
lead to complex conjugation of states in the string spectrum. From the
field theory analysis we know that the only possible anomalous
$\U{1}$ generators  are $q_b = \sum_J v_b^J H_{b}^{J}$ 
where $v_b^J$ is the
gauge  shift and $H_{b}^{J}$ is an element of the Cartan subalgebra of
the gauge group. In table \ref{tb:Z3shifts} have given the traces of
$q_b$ of the untwisted and twisted zero modes. In string theory we
probe the trace of these charges by calculating the expectation value
of $q_b = \sum_J v_b^J \, V_j^{bJ}$.

\begin{table}
{\small 
  \begin{center}
\renewcommand{\arraystretch}{1.5}
  \begin{tabular}{|l|c|l|c|c|}\hline
    {model}
      & {gauge shift}
& \multicolumn{1}{|c|}{gauge group}
& \multicolumn{2}{|c|}{$\U{1}$ trace} 
\\ 
& {$~~(v_1^I ~|~ v_2^I)$} & \multicolumn{1}{|c|}{$G $} 
& {$(q_1, q_2)_{un}$} & {$(q_1, q_2)_{tw}$}
      \\\hline
    $\E{8}$
      & $\frac{1}{3}\!\left(~0^8 ~~~~ ~~~~~ ~~|~~ 0^8 ~~~~~ ~~~~~ \right)$
        &    $~~~~~~ ~~~~~~ ~\, \E{8}  \times   \E{8}'$
& &       \\\hline
    $\E{6}$
      & $\frac{1}{3}\!\left(\mbox{-} 2,~1^2,~ 0^5 ~~|~~ 0^8 ~~~~~ ~~~~~ \right)$
       & $~~~ \E{6}\!\times\!\SU{3} \times \E{8}' ~~~~~~~ $
& &      \\\hline
    $\E{6}^2$
      & $\frac{1}{3}\!\left(\mbox{-}2,~1^2,~0^5 ~~|~~ \mbox{-}2,~1^2,~0^5\right)$
      & $~~~ \E{6}\!\times\!\SU{3} \times \E{6}'\!\times\!\SU{3}'$
& &      \\\hline
    $\E{7}$
      & $\frac{1}{3}\!\left(~0,~1^2,~0^5 ~~|~~ \mbox{-}2,~0^7 ~~~~~ \right)$
      &  $ ~~~~\,  \E{7}\!\times\!\U{1} \times
\SO{14}'\! \times\! \U{1}' \!\!$
& {$(6,2)$} & {$(10,\mbox{-}2)$}      \\\hline
    $\SU{9}$
     & $ \frac{1}{3}\! \left(\mbox{-}2,~1^4~,0^3 ~~|~~ \mbox{-}2,~0^7
~~~~~ \right) $
      & $~~~~~ ~~~~ \SU{9} \times \SO{14}'\!\times\!\U{1}' \!\!$
& {$(0,2)$} & {$(0,6)$}      \\\hline
    \end{tabular}
  \end{center}
}
  \caption{The defining gauge shifs $(v_1 | v_2)$ and the resulting
unbroken zero mode gauge groups are displayed for the five $\Intr_3$
orbifold models.  The last two columns give the zero mode traces of
the generators $q_b$ over the untwisted ($un$) and twisted ($tw$)
sectors, when applicable. 
}  
  \label{tb:Z3shifts}
\end{table}

All the classifying gauge shifts of table \ref{tb:Z3shifts} can be
represented as   
\equ{
v = \frac 13 \pmtrx{
1^{2 r^1_1} & \mbox{-}2^{r^2_1} & 0^{8-2 r^1_1 - r^2_1} 
& \Big| & 
1^{2 r^1_2} & \mbox{-}2^{r^2_2} & 0^{8-2 r^1_2 - r^2_2} 
}, 
}
for some integers $r_b^a$. 
This shift fulfills the constraints \eqref{consistency} when 
$r_1^1 + r_2^1 = r_1^2 + r_2^2$. We apply a similar
short--hand notation for products of $r^a_b$ theta functions, and write 
\equ{
\gvth\brkt{\ga}{\ga'}(\gn^{r^a_b}|\gt) = 
\prod_{I \in r^a_b} \gvth\brkt{\ga}{\ga'}(\gn_b^I|\gt),
}
identifying the index $r^a_b$ with the corresponding set of $1$'s and
$-2$'s. Similarly, $8 - 2r_b^1 -r_b^2$ denotes the set of $0$'s. With
this notation and the periodicities of the characteristics of the
theta functions \eqref{charperiod}, the holomorphic partition function
\eqref{HoloPartFun} reads 
\equ{
Z^{p,p'}_{t_a, t_a'}(\gn|\gt) = 
\frac {-1}{2\pi} 
\hat \get^{p,p'}_{t_a, t_a'} \, 
\frac 1{\get(\gt)^{15}} 
\prod_{a} \gvth\brkt{\frac{1-t_a}{2} - \frac {a}{3}}
{\frac{1-t_a'}{2} - \frac {p'}{3}}(\gn^{2r^1_a+ r^2_a}|\gt)
\gvth\brkt{\frac{1-t_a}{2}}
{\frac{1-t_a'}{2}}(\gn^{8 -2r^1_a- r^2_a}|\gt)
\gvth\brkt{\frac 12 - \frac{p}3}{\frac 12 - \frac{p'}3}^{-3}(0|\gt).
\labl{HoloPartFunSys}
}
with the modified phase factor 
\(
\hat\get^{p,p'}_{t_a,t_a'} = 
\tget^{p,p'}_{t_a,t_a'}
\, \exp\bigl({2 \gp i p' \bigl[  
\sum_a r_a^2 ( \frac {1-t_a}2 - \frac p3) + \frac p3 - \frac 12 
\bigr]}\bigr).
\)
The expectation value of this charge $q_b$ can be conveniently
computed in any particular sector as the derivative of the 
character--valued holomorphic partition function \eqref{HoloPartFun}.
This takes the form of the twisted fermionic correlator
\eqref{NormFermProp} 
\equ{
Q_b|^{p,p'}_{t_a, t_a'}(\gt) Z ^{p,p'}_{t_b, t_b'}(\gt) 
=  
\sum_J v^J_b \frac{\der}{\der \gn_b^J} 
 Z ^{p,p'}_{t_a, t_a'}(\gn|\gt) \Big|_{0} 
= 2(r_b^1 - r_b^2) 
\frac {\der}{\der \gn} 
\ln \gvth\brkt {\frac{1-t_a}{2} - \frac {p}{3}}
{\frac{1-t_a'}{2} - \frac {p'}{3}}(\gn|\gt) \Big|_{0}
Z ^{p,p'}_{t_a, t_a'}(0|\gt), 
\labl{Charges}
}
using the holomorphic partition function in the form
\eqref{HoloPartFunSys}.

\subsection*{Non--anomalous models}

From the expression \eqref{Charges} we immediately conclude that the
$\E{8}$, $\E{6}$ and $\E{6}^2$ models, defined by the gauge shifts
given in table \ref{tb:Z3shifts}, do
not have any local (and therefore integrated) tadpoles in string
theory, because they have the special property that $r_a^1 = r^2_a$
for both $a=1$ and $2$. This is quite a remarkable result since this is
not a statement concerning the zero modes (from a four or ten
dimensional point of view) of the string theory only, but is exact and 
based only on the gauge shifts, not on any particular property of the 
amplitudes themselves. This gives a direct string
confirmation of the field theory results presented in ref.\
\cite{GrootNibbelink:2003gb} based on the zero mode spectrum only. 
Of course, one could argue that this was to be expected since the four
dimensional gauge groups do not contain any $\U{1}$ factors, and
certainly not any anomalous $\U{1}$. By contrast, it should be noted
that, even though $\SU{9}$ is also a non--Abelian group, we cannot use
the same string argument to show that the trace of the $\U{1}$
generator for this group vanishes, since  table \ref{tb:Z3shifts}
shows that $r_a^1 \neq r^2_a$.

\subsection*{Anomalous models}

Next we move to the models that have an anomalous $\U{1}$ in their 
zero mode matter spectrum. To address the question, which states
contribute to the local tadpoles at one loop in string theory, it is
convenient to derive power series expansions  of the functions
$Q_s(\gt)$ in $q = \exp (2 \pi i\,\gt)$. The masses of the relevant
states are encoded as the power of $q$ in these expansions. In
particular, a negative power signals that tachyons give non--vanishing
effects. In table \ref{tb:Qcontr} we only quote the leading order results; 
they already give an interesting insight in the contributing
states, as we now discuss in detail for both anomalous models
individually.

\begin{table}
\[
\renewcommand{\arraystretch}{1.25}
\arry{| l  r | c | c | c | c |}{
\hline 
\text{model} & \text{charge} & \text{~~$u$--sector~~} & 
\text{~~$t$--sector~~} & \text{$d_+$--sector} & 
 \text{$d_-$--sector} 
\\ \hline 
\SU{9} & Q^1_s(\gt) & 
 0 & 0  & 0  &  0
\\[1ex] 
& Q^2_s(\gt)  & 
2 & 2 & 2 & 2  
\\[1ex] 
\hline & & & & & 
\\[-3ex]
\E{7} &  Q^1_s(\gt) & 
3 + \ldots & 
\frac 19 q^{-\frac 13} + \frac 53 + \ldots  &
\frac 19 e^{i \frac {4\pi}3} q^{-\frac 13} + \frac 53 + \ldots  &  
\frac 19 e^{i \frac {2\pi}3} q^{-\frac 13} + \frac 53 + \ldots 
\\[1ex] 
& Q^2_s(\gt) & 
2 + \ldots & 
\frac 29 q^{-\frac 13} - \frac 23 + \ldots & 
\frac 29 e^{i \frac {4\pi}3} q^{-\frac 13} - \frac 23 + \ldots  &  
\frac 29 e^{i \frac {2\pi}3} q^{-\frac 13} - \frac 23 + \ldots    
\\[1ex]
\hline 
}
\]
\caption{The charges $Q^b_s(\gt)$ as functions of 
$q = \exp(2 \pi i\, \gt)$ for the two anomalous models
are displayed for the four different sectors $u, t, d_+$ and $d_-$
defined in section \ref{sc:twProp}. The results for the $\SU{9}$ model
are exact, while for the $\E{7}$ model the dots indicates that all
massive string states are neglected here. The tachyonic contribution
within the various twisted sectors ($t, d_\pm$) cancel among
themselves.  
\labl{tb:Qcontr}}
\end{table}

We begin with the situation in the $\SU{9}$ model. The absence of
the dots is meant to indicate that results for the expressions for
$Q_s(\gt)$ for the various sectors $s$ are exact. Since there are only
constants presents, only massless string modes contribute to the gauge
field tadpole. This suggests for this model the effective field theory
approach, describing only the zero modes, takes account of all
contributions. Moreover, since the zero mode gauge group is 
$\SU{9} \times \SO{14}'\!\times\!\U{1}'$ and only zero modes
contribute, it comes as no surprise that the traces of $Q^1_s$ vanish
for all sector $s$ separately. The
traces of $Q^2_s$ are equal for all four sectors. Moreover, the sum of
charges of the  untwisted sector ($un = u$) and the twisted sectors 
($tw = t, d_+$ and $d_-$) are equal to $2$ and $6$, respectively, see
table \ref{tb:Qcontr}, which agrees with results obtained in ref.\ 
\cite{GrootNibbelink:2003gb,Gmeiner:2002es} quoted in the last two
columns of table \ref{tb:Z3shifts}.

These results of the $\SU{9}$ model are in sharp contrast to the
situation in the $\E{7}$ model as the bottom part of table
\ref{tb:Qcontr} demonstrates. Their only similarity 
is that the $\E{7}$ model also passes the cross check
that the sums of the zero mode charges of tables \ref{tb:Z3shifts} and
\ref{tb:Qcontr} agree. But in
addition to the zero modes, whole towers of massive string states
contribute to the gauge field tadpoles as the dots indicate. 
However, it can be shown that for the integrated
tadpole, which simply sums up the contributions of all sectors, these
massive excitations cancel out. More surprisingly, there are also
tachyonic contributions to the different twisted sectors. It is not
difficult to see from the table that also they cancel among themselves
when one considers the integrated gauge field tadpole.

The situation in the $\E{7}$ model may be summarized as follows: 
At the four dimensional zero mode level, only the first $\U{1}$,
generated by $q_1$, is anomalous. But locally we see that, because the
various sectors have different profiles over the internal dimensions,
both $\U{1}$ are anomalous and both tachyonic and massive states  
contribute to them. The conventional effective field theory
description of this model is able to make the distinction between the
momentum profiles due to the massless untwisted and the twisted 
states. Field theory does not determines its own spectrum, therefore 
it lacks the ability to predict the extra contributions of
tachyonic (and massive string) states. In our next paper we will
investigate how important of these massive and tachyonic states are
for the final profile of the tadpoles.

In closing this section, we would like to remark that we have also
computed the local tadpole in the $\SO{32}$ string. In this case the
standard embedding ($v = ( 1^2, \mbox{-}2, 0^{13}/3)$) gives rise to an
anomalous $\U{1}$ generated by $q = w_I H_I^1$ with 
$w = (1^3, 0^{13})/3$. 
The properties of the local gauge field tadpole are
very similar to the $\SU{9}$ model of the $\E{8}\times \E{8}'$ string:
Also for the $\SO{32}$ string we found that only 
the zero modes have non--vanishing profiles over the extra
dimensions. The integrated tadpole gives results consistent with 
the results quoted in ref.\ \cite{Atick:1987gy}

%% file: Sconcl.tex
\section{Conclusions}
\labl{sc:concl}

We have computed local gauge field tadpoles in heterotic 
$\E{8}\times \E{8}'$ strings on the non--compact orbifold
$\Cplx^3/\Intr_3$. This calculation confirms recent field theoretical 
calculation of such tadpoles, but at the same time extends these
results in various interesting and surprising directions. In detail,
our findings are the following:

The shape of these tadpoles are governed by the propagators of the
twisted coordinate fields on the string world sheet: Their expectation
values $\langle X^\ui X^i \rangle(\gt)$ determine the widths of
Gaussian momentum distributions on the orbifold. The propagators can
be classified according to their orbifold boundary conditions; the
corresponding sectors give rise to Gaussian distributions of various 
widths. The momentum distribution of the total tadpole is obtained by
integrating these Gaussians over the fundamental domain of the
Teichm\"uller parameter of the one loop world sheet torus. This means
that the tadpole in coordinate space is not a simple Gaussian
distribution, but a sum of Gaussians weighted by the respective
partition functions which is integrated over the fundamental domain.

The propagators for the twisted coordinate fields $X^i$ and $X^\ui$ at
zero separation are determined up to a single universal normal ordering
constant associated with the subtraction of the logarithmic
singularity of the untwisted correlators at zero separation. Conventional 
wisdom suggests that this normal ordering constant should be
irrelevant as it drops out of on--shell string amplitudes. However,
since the gauge field tadpole is necessarily an off--shell quantity,
the final expression does depend on this normal ordering
constant. Moreover, if this constant becomes too small, the coordinate
space expression for the tadpole becomes ill--defined. This determines
a lower bound for the value of this normal ordering constant.

We found that all eight bosonic propagators for the different twist
sectors on the torus world sheet can be expressed in terms of four 
fundamental correlators that have different dependence on the modular
parameter $\gt$. This implies that both four dimensional untwisted and
twisted sectors have distinct non--trivial profiles over the extra
dimensions. In the field theory discussion of string orbifold models
one usually assumes that the twisted sector states live exactly at the
orbifold fixed points. Our calculation suggests that this is an
approximation insensitive to string physics in which the twisted states
are spread out around the orbifold singularity. One of the objectives
of our follow--up paper \cite{GNL_II} is to investigate just how crude
the conventional field theory approximation really is; this leads some
suggestions how fixed point states could be treated in field
theoretical models.

Finally, we have investigated which states contribute in the different
sectors to the momentum or coordinate profiles of the gauge field
tadpoles. The anomalous $\Intr_3$ model containing the gauge  
group $\SU{9}$ complies with the expectation that only the zero modes
participate. However, quite surprisingly, we found that for the other
anomalous model massive and even tachyonic string states
contribute. We emphasize that this is not in contradiction with
previous results in the literature for the zero 
mode $D$--term: We have verified that on--shell, which for a tadpole
means that $k_i =0$, the contributions from all tachyonic and massive
states vanish.  However, for generic $k_i \neq 0$ the effects of these
states do not entirely cancel out when the four orbifold sectors are
combined. In our next paper we explore the significance of these
contributions further numerically.

\section*{Acknowledgements}

We would like to thank O.\ Lechtenfeld, R.\ Myers,   
A.N.\ Schellekens and M.\ Walter for useful discussions during 
various stages of this project. We are also grateful to M.\ Serone for
helping us resolve a problem with modular invariance in a previous
version of the manuscript. The work of S.G.N.\ was supported partly
by DOE grant DE--FG02--94ER--40823. The work of M.L.\ was 
partially supported by NSERC Canada. S.G.N.\ would like to thank the
University of British Columbia for their hospitality during which this
project got started.  M.L.\ would like to thank the
University of Minnesota for their hospitality during part of the
preparation of this work.

%% file: Swstorus.tex
\section{World sheet torus theories}
\labl{sc:wstorus}

In this appendix we collect some results concerning fermionic and
bosonic conformal field theories on the torus world sheet. Most
results are well--known, but for completeness and to fixed our
notations and conventions we review them here. More pedagogical
discussions can be 
found in ref.\ \cite{gsw_1,gsw_2,Ginsparg:1988ui,pol_1,pol_2,
Kiritsis:1997hj}. Many of the properties of conformal
field theories can be conveniently encoded by theta and related
functions. We have summarized their properties in our conventions 
in appendix \ref{sc:theta}.

The complex world sheet world sheet torus coordinate $\gs$ satisfies
the periodicities $\gs \sim \gs +1$ and $\gs \sim \gs + \gt$, where
$\gt = \gt_1 + i \gt_2$ is the modular parameter labeling conformally
distinct tori. The volume of the torus is $\int \d^2\gs = 2\, \gt_2$. 
We define the functions 
\equ{
\gF\brkt{\ga}{\gb}(\gs,\bgs| \gt) = 
\exp\Bigl(2\pi i\, 
\frac { (\gt \ga + \gb) \bgs - (\bgt \ga + \gb) \gs}{\bgt - \gt}
\Bigr), 
\labl{modes}
} 
that have the following periodicity properties on the world sheet
torus 
\equ{
\gF\brkt{\ga}{\gb}(\gs + 1) = e^{- 2\pi i\,\ga}\, \gF\brkt{\ga}{\gb}(\gs),
\qquad 
\gF\brkt{\ga}{\gb}(\gs + \gt) = e^{+2\pi i\,\gb}\, \gF\brkt{\ga}{\gb}(\gs),
\labl{modefun}
} 
Here we have suppressed part of the arguments of these functions 
for brevity.  Under modular transformations we have the transformations
\equ{
\gF\brkt{\ga}{\gb}(\gs | \gt+1) = \gF\brkt{\ga}{\gb+\ga}(\gs | \gt), 
\qquad 
\gF\brkt{\ga}{\gb}\Bigl( \frac \gs\gt \Bigr|\Bigr. \frac{\mbox{-}1}{\gt} \Bigr)
= 
\gF\brkt{+\gb}{-\ga}(\gs | \gt). 
\labl{modularP}
}

\subsection{Complex fermion}
\labl{sc:fermion}

Let $\gl$ be a complex fermion on the world sheet torus which obeys
the boundary conditions 
\equ{
\gl(\gs+1) =  e^{- 2\pi i\,\ga}\, \gl(\gs),
\qquad 
\gl(\gs+\gt) =  e^{+2\pi i\,\gb}\, \gl(\gs), 
}
and the complex conjugate boundary conditions for $\bgl$. The mode
expansion for $\gl$ can be expressed using \eqref{modefun} as 
\equ{
\gl(\gs) = \sum_{m,n} \gF\brkt{m+\ga}{n\,\, +\gb}(\gs) \, c^m_n, 
}
where the sum is over all integers $m, n\in \Intr$. The (quantized)
harmonic oscillators are denoted by $c^m_n$. One can define free
left--moving (holomorphic) theory for these boundaries by the actions  
\equ{
S_{\gl} \brkt{\ga}{\gb}(\gm|\gt) = -\frac{1}{2\gp} 
\int \d^2\gs\, \Bigl( 
\bgl\, \bder\, \gl  + \frac {2 \pi i}{\bgt - \gt} \gm \, \bgl \gl
\Bigr) 
= \sum_{m,n} 
\bigl[ \gt(m+\ga) + n + \gb + \gm \bigr] \,  \bc^{-m}_{-n} c^m_n. 
}
Here we have introduced a source term $\gm$, for later convenience. 
We note that because  the spectrum of the right--moving
(anti--holomorphic) theories (with $S_\gl$) is conjugate to the
spectrum of the left--moving theory, we need 
only give the expressions for
the holomorphic theory ($S_{\gps}$) here.
The (character valued) partition function \cite{Schellekens:1987xh} 
is given  in terms of the theta function \eqref{thetaGen} and the
Dedekind $\get$--function \eqref{Dedekind} by 
\equ{
Z_{\gl}\brkt{\ga}{\gb}(\gm | \gt) 
= 
e^{2\pi i\, \ga \bigl( \frac {\gt}{2} \ga  + \gb + \gm - \frac 12 \bigr)}\, 
\int \prod_{m,n} \d c^m_n \d \bc^{-m}_{-n} \, 
e^{- S_{\gl}\brkt{\ga}{\gb}(\gm | \gt)}
= 
\frac{ \gvth\brkt{\frac 12- \ga}{\frac 12 - \gb}(\gm |\gt) }{\get(\gt) }. 
\labl{eq:freefermpf}
} 
This result is obtained by
discarding (infinite) constant factors, $\gz$--function regularization
techniques \cite{Ginsparg:1988ui} and the product expansion of the
theta function $\gvth\brkt{\ga}{\gb}$ given in \eqref{prodtheta}. 
The phase factor in front of the path integral has been chosen such
that the final result can be written in terms of the theta function
with characteristics $\frac 12 -\ga$ and $\frac 12 - \gb$.

Using standard field theory techniques the propagator (for $\gm = 0$) 
can be determined 
\equ{
\tgD_\gl\brkt{\ga}{\gb}(\gs|\gt) 
= - 
\sum_{m,n} \frac{ \gF\brkt{m+\ga}{n\,\,+\gb}(- \gs|\gt) } 
{ \gt(m+\ga) + n + \gb } 
= -
 \frac{ 
 \gvth_1'(0|\gt)\, \gvth\brkt{\frac 12 - \ga}{\frac 12 -\gb}(\gs|\gt)}
{ \gvth_1(\gs|\gt)\, \gvth\brkt{\frac 12 - \ga}{\frac 12 -\gb}(0|\gt)}.
\labl{eq:ffprop}
}
As usual the propagator only depends on the relative 
world--sheet separation. We use the notation $\tgD$ to
denote two--point correlation functions at non--vanishing separation
$\gs$. The conformally normal ordered expression of the propagator is
defined by \cite{pol_1,pol_2}
\equ{
\gD_\gl \brkt{\ga}{\gb}(\gt) = 
 \lim\limits_{\gs \ra 0} \, 
\langle  \bgl(\gs) \gl(0) \rangle\brkt{\ga}{\gb}(\gt)  
+
 \frac 1{\gs}, 
\labl{DiffEqProp}
}
in the limit of zero separation $\gs \ra 0$. Using the theta function
expression for the propagator \eqref{eq:ffprop}, this becomes 
\equ{
\gD_\gl \brkt{\ga}{\gb}(\gt) 
= -
\frac{\gvth\brkt{\frac 12 - \ga}{\frac 12 -\gb}'(0|\gt)}
{\gvth\brkt{\frac 12 - \ga}{\frac 12 -\gb}(0|\gt)} 
= - 
\left. \frac {\der}{\der \gm}\, \ln Z_{\gl}\brkt{\ga}{\gb}(\gm | \gt)  
\right|_{\gm = 0}. 
\labl{NormFermProp}
}
The second expression gives an alternative way to derive this
expectation value, using the character valued partition function
\eqref{eq:freefermpf}.

\subsection{Complex boson} 
\labl{sc:boson}

We consider a complex boson $X$ on the string world sheet with
boundary conditions 
\equ{
X(\gs + 1) = e^{-2 \pi i\, \ga} X(\gs), 
\qquad 
X(\gs + \gt) = e^{+2 \pi i\, \gb} X(\gs). 
\labl{boundBoson}
}
As for the fermions, the mode expansion is
\equ{
X(\gs) = \sum_{m,n} \gF\brkt{m+\ga}{n\,\, +\gb}(\gs) \, a^m_n,
}
and the dynamics of the boson are described by the free action 
\equ{
S_X\brkt{\ga}{\gb}(\gt)  = - \frac 1{2\pi} \int \d^2\gs
\Bigl( \der \bar X \, \bder\, X + \bder \bar X\, \der X \Bigr)
= 2 \gp\,  \sum_{m,n} \frac{ | \gt(m+\ga) + n + \gb |^2}{2 \gt_2} \, 
\baa^{-m}_{-n} a^m_n. 
} 
The resulting partition function takes the form 
\equ{
Z_X \brkt{\ga}{\gb}(\gt) = 
\Bigl| 
e^{2\pi i\, \bigl\{ \frac 12 \ga^2 \gt + \ga(\gb - \frac 12) \bigr\}} 
\Bigr|^2
\int \prod_{m,n} \d a^m_n \d \bar a^{-m}_{-n} \, 
e^{-S_X \brkt{\ga}{\gb}(\gt) }
=
\frac{ \bigl| \get(\gt) \big|^2} 
{ \bigl | \gvth\brkt{\frac 12 - \ga}{\frac 12 -\gb} ( 0 | \gt )\big|^2}.
\labl{eq:freebospf}
}
Here we have used the same phase factor as in \eqref{eq:freefermpf}. 
As for the fermions, the formal expression for the propagator reads 
\equ{
\tgD_X\brkt{\ga}{\gb}(\gs|\gt) 
= -
\frac 1{2\pi} \sum_{m,n} 
\frac { 2\gt_2}{|\gt(m+\ga) + n + \gb|^2} 
\gF\brkt{m+\ga}{n\, +\gb}(- \gs| \gr).
\labl{eq:fbprop}
}
For any of these boundary conditions (which we do not write explicitly
here) one can derive that 
\equ{
\int \cD X\,  e^{
i k\, \bX(\gs) + i l\, \bder \bX(\gs)
+ i \bar k\, X(\gs') + i \bar l\, \bder X(\gs')
}  =
e^{- \bar k k \, \tgD_X + k \bar l\, \bder \tgD_X - \bar k l \, 
\bder \tgD_X
+ \bar l l\, \bder\bder \tgD_X}(\gs'-\gs)
\labl{ExpCorr}
}
for arbitrary $k, \bar k, l$ and $\bar l$.

The properties of the twisted propagators are of central importance to
our work, and are, therefore, discussed in section \ref{sc:twProp} of
our main discussion.  As these propagators can be expressed in terms
of the untwisted bosonic propagator, with $\ga = \gb = 0$,  we review
its properties in this appendix. 
The formal series expansion of the untwisted propagator reads 
\equ{
\tgD(\gs|\gt) 
= -
\frac 1{2\gp} \sum_{m,n}' \frac{ 2 \gt_2} { | \gt m + n|^2 } 
 \, \gF\brkt{m}{n}(- \gs|\gt),
\labl{untCorr}
}
where the prime on the sum indicates that the sum is over all integers
with $(m,n) \neq (0,0)$. It follows that the regularized correlator is the
solution of 
\equ{
\bder \der\, \tgD(\gs|\gt) 
= 
2 \gp\,  \Bigl( \gd^2(\gs) + \frac 1{2\gt_2} \Bigr),
}
which is required to be modular invariant and periodic. Here we have
chosen the same normalization for the bosonic propagator as for the
fermionic propagator with respect to the delta function $\gd^2(\gs)$
in the defining differential equation \eqref{DiffEqProp}. This
correlator can be expressed in terms of theta functions as  
\equ{
\tgD(\gs|\gt) 
= -
\ln \tilde G(\gs| \gt), 
\qquad 
\tilde G(\gs | \gt) = 2\gt_2\, 
e^{-2\pi\, \frac {\gs_2^2}{\gt_2}} \, 
\Bigl| \frac{\gvth_1(\gs| \gt)}{\gvth_1'(0|\gt)} \Bigr|^2.
\labl{UnProp}
} 
Notice that this fixes $\tgD$ up to an additive constant. (Fortunately,
for the determination of the twisted propagator this undetermined
constant is irrelevant, see section \ref{sc:twProp}.) 
Finally, the normal ordered untwisted propagator at zero separation is
given by  
\equ{
\gD(\gt) = \lim\limits_{\gs \ra 0}  
 \langle : \! \bar X (\gs) X(0) \! : \rangle\brkt{0}{0}(\gt) = 
\tgD(\gs| \gt) + \ln {|\gs|^2} + \tilde c = - \ln  (2\gt_2) + \tilde c,
\labl{NormUnProp}
} 
with $\tilde c$ a normal ordering constant.


%% file: Stheta.tex
\section{Theta functions}
\labl{sc:theta}

The genus one theta function is defined by 
\equ{
\gvth\brkt{\ga}{\gb} (\gs | \gt) 
= 
\sum_{n\in \Intr} \, q^{\frac 12 (n - \ga)^2} \, 
e^{2\gp i\, (\gs - \gb) (n - \ga)}, 
\qquad 
q = e^{2\gp i\, \gt}. 
\labl{thetaGen}
}
In a product representation it takes the form: 
\equ{
\gvth\brkt{\ga}{\gb}(\gs | \gt) = 
e^{- 2 \pi i \ga( \gs - \gb)}
\, q^{\frac 12 \ga^2} \, 
\prod_{n \geq 1} \Bigl\{
\bigl( 1 - q^n \bigr) 
\prod_{s= \pm} 
\bigl( 1 + e^{2 \pi i s(\gs - \gb)}\, q^{n - \frac 12 - s \ga} \bigr) 
\Bigr\}.
\labl{prodtheta}
}
The arguments of the theta functions are periodic in the sense that 
\equ{
\gvth\brkt{\ga+1}{\gb}(\gs | \gt) = \gvth\brkt{\ga}{\gb}(\gs | \gt), 
\qquad 
\gvth\brkt{\ga}{\gb+1}(\gs | \gt) = e^{2\pi i\, \ga}\, \gvth\brkt{\ga}{\gb}(\gs | \gt).
\labl{charperiod}
}
Modular transformations have the following effect on the theta
functions
\equ{
\gvth\brkt{\ga}{\gb}(\gs|\gt+1) = e^{-\gp i\, \ga(\ga+1)}\, 
\gvth\brkt{\ga}{\gb+\ga+\frac 12}(\gs|\gt), 
\quad 
\gvth\brkt{\ga}{\gb}
\Bigl(\frac {\gs}{\gt}\Bigr| \Bigl. \frac{\mbox{-}1}{\gt}\Bigr) 
= 
\sqrt{- i \gt} \, e^{2 \gp i\,\bigl\{ \frac {\gs^2}{2\gt} + \ga\gb \bigr\}} 
\gvth\brkt{+\gb}{-\ga}(\gs| \gt). 
\labl{modulartheta} 
}
The periodicities of the theta functions read 
\equ{
\gvth\brkt{\ga}{\gb}(\gs+1|\gt) = e^{-2\gp i\, \ga}
\, \gvth\brkt{\ga}{\gb}(\gs|\gt), 
\qquad
\gvth\brkt{\ga}{\gb}(\gs+\gt |\gt) = e^{2\gp i\, (\gb - \gs - \frac 12 \gt)}
\, \gvth\brkt{\ga}{\gb}(\gs|\gt). 
\labl{periodtheta} 
}
An often used notation is $\gvth_1 = \gvth\brkt{1/2}{1/2}$, 
$\gvth_2 = \gvth\brkt{1/2}{0}$,  $\gvth_3 = \gvth\brkt{0}{0}$,  
and $\gvth_4 = \gvth\brkt{0}{1/2}$. 
Another important modular function is the Dedekind $\get$--function 
\equ{
\get(\gt) = q^{\frac 1{24}} 
\prod_{n \geq 1} \bigl( 1 - q^n \bigr), 
\qquad 
\gvth_1'(0|\gt) = 2\gp\, \bigl(\get(\gt) \bigr)^3, 
\labl{Dedekind}
}
where the prime $'$ denotes differentiation by the first argument
of $\gvth_1$. The modular transformation properties of the Dedekind
function take the form 
\equ{
\get(\gt + 1) = e^{2 \pi i \frac 1{24}} \get(\gt), 
\qquad 
\get(\frac {-1}{\gt}) = \sqrt{-i \gt} \get(\gt).
}
The Riemann's identity reads 
\equ{
\sum_{s,s'=0,1} e^{- \gp i\, s s'} \prod_{i=0}^3
\gvth\brkt{\frac 12 - \ga_i - \frac {s}2}{\frac 12 - \gb_i - \frac {s'}2}
(\gm_i) 
= 2 \,  \prod_{i=0}^3
\gvth\brkt{\frac 12 + \ga_i}{\frac 12 + \gb_i}
(\tgm_i)
= 2 \,  \prod_{i=0}^3
\gvth\brkt{\frac 12 - \ga_i}{\frac 12 - \gb_i}
(-\tgm_i), 
\labl{RiemannSimple}
}
with $\tgm_i = \frac 12 \sum_k \gm_k - \gm_i$ and  
\(
\frac 12 \sum_i \ga_i =\frac 12 \sum_i \gb_i = 0 \mod 1. 
\)